# Weaves: A Novel Direct Code Execution Interface for Parallel High Performance Scientific Codes

Srinidhi Varadarajan, Joy Mukherjee, and Naren Ramakrishnan
Department of Computer Science, Virginia Tech, Blacksburg, VA 24061
Email: {srinidhi,jmukherj,naren}@vt.edu

**Abstract**

Scientific codes are increasingly being used in compositional settings, especially problem solving environments (PSEs). Typical compositional modeling frameworks require significant buy-in, in the form of commitment to a particular style of programming (e.g., distributed object components). While this solution is feasible for newer generations of component-based scientific codes, large legacy code bases present a veritable software engineering nightmare. We introduce Weaves – a novel framework that enables modeling, composition, direct code execution, performance characterization, adaptation, and control of unmodified high performance scientific codes. Weaves is an efficient generalized framework for parallel compositional modeling that is a proper superset of the threads and processes models of programming. In this paper, our focus is on the transparent code execution interface enabled by Weaves. We identify design constraints, their impact on implementation alternatives, configuration scenarios, and present results from a prototype implementation on Intel x86 architectures.

## 1 Introduction

The past decade has witnessed increasing commoditization of scientific computing codes, leading to the prevailing practice of compositional software development. The ability to combine representations for different aspects of a scientific computation to create a representation for the computation as a whole is now considered central to high-level problem solving environments (PSEs), especially for the grid [Natrajan et al., 2002; Ramakrishnan et al., 2002]. Many PSEs provide interfaces where the scientist describes data-flow relationships between codes in the form of a graphical network and the PSE manages the details of composing the application represented by the network. Such compositional modeling can be employed at all stages of computational simulation - model specification, model execution, and model analysis. In this paper, we are interested in the former two stages; connections to model analysis (performance modeling of scientific codes; e.g., [Browne et al., 2000]) will be made later in the paper. The common solutions for model specification and execution harness technologies like encapsulation and distributed OO [Gannon and Grimshaw, 1998], parallel programming primitives [Foster, 1996], agent-based composition [Drashansky et.al., 1999], and markup languages.

Concomitantly, there is a renewed emphasis on accommodating application-specific considerations in compositional modeling [Adve and Sakellariou, 2000; Decker and Wylie, 1997]. The motivation here is to provide sophisticated performance modeling [Adve et al., 2000] and/or support adaptive control [Adve et al., 2002] of scientific computing applications; for example, algorithm selection, runtime steering, and checkpointing can be better provided with knowledge of the underlying application.

It is difficult to create an integrated modeling framework that seamlessly supports model specification and execution and, at the same time, flexibly allows the incorporation of application-specific considerations. Traditional approaches require significant *buy-in*, in the form of commitment to a particular style of programming or an implementation technology (e.g., distributed object components). This is a serious hurdle for large, legacy code bases, which cannot be easily ported to a new modeling framework. The primary requirement here is for a *transparent* framework that supports modeling, composition, execution, performance characterization, adaptation, and control of high performance scientific codes. In this paper, we present **Weaves**, which provides a direct code execution interface to this functionality.

## 2 Direct Code Execution Interfaces to Scientific Codes

Our use of the term *direct code execution* (**DCE**) is in the context of executing *unmodified* scientific codes. Creating a direct code execution framework presents several interesting research challenges:

1. *Scalability*: The scale of the direct code execution system may be vastly different from the intended target of the scientific code base. For instance, a 10-processor cluster may be used as the basis to create a performance model of a scientific code base that requires 1000 processors. Handling this difference in scale requires a framework that can efficiently create a virtual machine (**VM**) abstraction and map physical processors in the DCE system onto the virtual machines.

2. *Transparency*: The implementation of the DCE framework should be transparent to the programming model used by the application. For instance, a direct code execution framework that executes unmodified applications cannot make any assumptions about the thread safety or reentrancy capabilities of the application.

3. *Representational Adequacy*: A DCE framework must support compositional development and execution of scientific codes; it should thus be powerful enough to represent the interaction between the various components of the original scientific application.

The two major direct code execution models today are the threads and processes models. In the threads model, direct code execution is achieved by running multiple threads through the same code. Since each thread has an independent control flow, this model creates multiple virtual machines, each of which executes one instance of the application. *The virtual machine abstraction created by the threads model is correct as long as the application being modeled does not use global state variables.* Each VM also exposes an emulated communication interface to allow the various threads to exchange messages. This approach is used by MPI-SIM [Prakash and Bagrodia, 1998], which uses a threaded model for direct code execution of applications that use MPI for communication.

In the processes model, direct code execution is achieved by executing (*fork* in UNIX systems) multiple independent instances of the same application. Since each process operates in its own virtual address space, with its own control flow, this model creates multiple virtual machines. The virtual machines communicate with each other through an emulated message passing interface, which in turn uses some form of inter-process communication (**IPC**) in a single processor system or a combination of IPC and message passing in a multiprocessor system.

We argue that neither the threads model nor the processes model is powerful enough to satisfy the scalability, transparency, and representation requirements from a DCE standpoint. We will elucidate this through case studies of a few sample applications.

**Example 1 (No Global State):** The first sample application is Sweep3D, a benchmark for 3D discrete ordinates neutron transport [Koch et al., 1992]. Available in FORTRAN in the public domain, Sweep3D forms the kernel of an ASCI application and involves solving a group of time-independent neutron particle transport equations on an XYZ Cartesian cube. The code uses a logical discretization of the 3D geometry, taking care to ensure that physical symmetries are not distorted, angular dependencies are preserved, and derivatives w.r.t. the angular coordinates are maintained. Sweep3D has been the focus of many important performance modeling efforts, most notably [Adve et al., 2000].

The application is written in Fortran 77, with dynamic array extensions. The main characteristic of Sweep3D is that it uses no global state. Since the application only relies on local state, multiple instantiations of local state are enough to create a VM abstraction. This characteristic makes Sweep3D inherently thread-safe, which enables its modeling by either the threads or processes models.

**Example 2 (Independent Global State):** Our second application is an interesting communication performance benchmark proposed by the editor of CiSE magazine [Sullivan, 1997]:

> "On each of 2048 processors, generate two random numbers *n1* and *n2*. Send *n2* copies of "hello world" to processor *n1*. Then each processor counts the number of instances of "hello, world" it has received. Do this several million times.
>
> Here, *n1* is bounded between zero and the total number of processors and *n2* is the parameter to vary as machines get better. This benchmark is interesting, because it really has no single-processor version. Hence, it can be used only to compare parallel machines with parallel machines. The data motion is typical of forms of parallel Monte Carlo. (Of course, except for the timings, the numbers are not very fascinating.)"

The main characteristic of this benchmark is that it involves asynchronous communication. Since the message receivers do not know the number of messages they will receive *a priori*, they cannot use any form of synchronous communication. This includes both blocking as well as non-blocking synchronous communication primitives. The only solution is asynchronous communication based on callback functions – similar to *hrecv()* calls in Paragon™ NX.

Asynchronous (or interrupt driven) communication inherently requires access to global state. In a typical sequential program, the use of global variables can be avoided by elaborate function parameterization, since the flow of control is known *a priori*. In codes based on callback functions – such as event driven codes - the sequence of callback function invocations is not known *a priori*, and hence the callback function cannot be parameterized completely. When the callback function is called – say on the receipt of a message - it has two possible options. It can either perform its actions at invocation or queue the message for later processing. The parameters to the callback function are generic, since the external agent invoking the callback function is not aware of the state required by the callback function. Hence, to perform any action that modifies program state, the state itself has to be globally scoped. If the callback function queues the message for later processing, the queue data structure has to be global. In either case, asynchronous communication requires access to global state.

To see how this relates to Sullivan's benchmark, note that message receivers are not aware of *n2*, the number of messages they will receive. To implement the benchmark, the main control flow specifies a callback function, which is automatically called when a message is received. The callback function updates a global count variable, which effectively counts the number of messages received at each receiver. Interestingly, standards compliant MPI 1.0, 1.1, 1.2 implementations do not support callback based asynchronous communication and hence cannot be used to implement Sullivan's benchmark. In [Skjellum et al., 1994], a simple extension to MPI - based on threads – is proposed to implement asynchronous communication. However, the extension assumes that the MPI libraries used are thread-safe. (In the strictest case, the application also has to be thread-safe, although this is not required for Sullivan's benchmark). While MPI 2 is heading in this direction, current implementations of MPI are not necessarily thread-safe.

The above example leads to two observations. First, asynchronous communication requires global variables, which violates thread safety assumptions. Second, Sullivan's benchmark has a very interesting property. The global variable count has to be independent for each instantiation of the application. This clearly rules out the threads model as a basis for a DCE framework. This benchmark can only be implemented by the processes model.

**Example 3 (Selectively Shared Global State):** Our final application involves the idea of *collaborating PDE solvers* [Drashansky et al., 1999] for solving heterogeneous multi-physics problems. For instance, simulating a gas turbine requires combining models for heat flows (throughout the engine), stresses (in the moving parts), fluid flows (for gases in the combustor), and combustion (in the engine cylinder). Each of these models can be described by an ODE/PDE with various formulations for the geometry, operator, and boundary conditions. The basic idea here is to replace the original multi-physics problem by a set of smaller simulation problems (on simple geometries) that need to be solved *simultaneously* while satisfying a set of interface conditions. The mathematical basis of this idea is the interface relaxation approach to support a network of interacting PDE solvers [McFaddin and Rice, 1992; see Fig. 1].

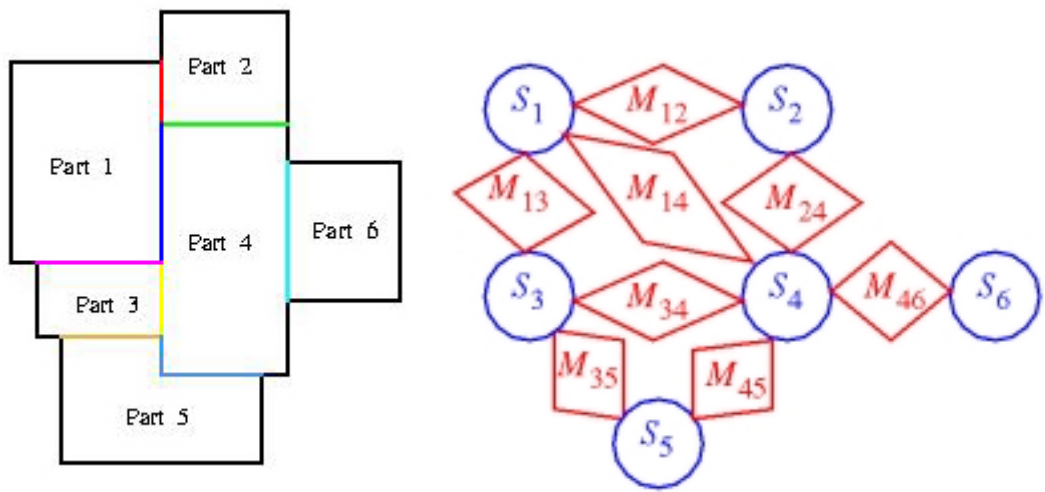

**Figure 1:** (left) Multi-physics problem with six subdomains with different PDEs. (right) A network of collaborating solvers (S) and mediators (M) to solve the PDE problem. Each mediator is responsible for agreement along one of the interfaces (colored lines).

Mathematical modeling of the multi-physics problem distinguishes between *solvers* and *mediators*. A PDE solver is instantiated for each of the simpler simulation problems and a mediator is instantiated for every interface, to facilitate collaboration between the solvers. The mediators are responsible for ensuring that the solutions (obtained from the solvers) match properly at the interfaces. The term "match properly" is defined by the physics if the interface is where the physics changes or is defined mathematically (e.g., the solutions should join smoothly at the interface and have continuous derivatives).

This application requires selective sharing of global state – no sharing of global state between the solvers and shared global state within the mediator. To support such selective sharing, we need mechanisms to separate as well as recombine global state. Since the threads model shares all global state, it cannot support arbitrary state separation *without code modification*. The processes model separates all global state, shifting the emphasis to mechanisms for recombining state. Intuitively, this is an easier problem to solve than separating global state.

Recombination of global state can be achieved through an agents model. It is thus not surprising that the collaborative PDE solvers problem has been approached using agent technology [Drashansky et al., 1999]. The critical observation here is that messages in agent technology are a powerful code-neutral abstraction for parameter passing and procedure invocation. Effectively, messages between agents are used to recombine state. For instance, two solver processes are used to separate their global state. To recombine state, the solver processes communicate with a single mediator process, whose state is a function of the messages received from the solvers.

*Our emphasis here is on a single programming model that can transparently support the above three classes of applications.* Note that the requirement of the collaborative PDE solvers application – arbitrary state sharing – is a generalization of the previous two examples. The current solution based on agents has used message passing as an indirect representation of procedural invocations, for arbitrary state sharing. However, message passing between processes does not satisfy the transparency and scalability requirements of a DCE framework. What we need is a programming model that supports arbitrary state sharing within a traditional procedural framework.

## 3   Weaves: A Compositional DCE Framework

Let us revisit the collaborating PDE solvers application from the context of the threads and processes models. We will illustrate the need for Weaves by a series of examples culminating in a DCE framework that can support the compositional model shown in Figure 2. For simplicity, assume that $S_1$ and $S_2$ are two instantiations of the PDE algorithm $A_1$ and $S_3$ and $S_4$ are two instantiations of a different PDE algorithm $(A_2)$.[1] From a scalability perspective, the DCE framework should operate within a single process.

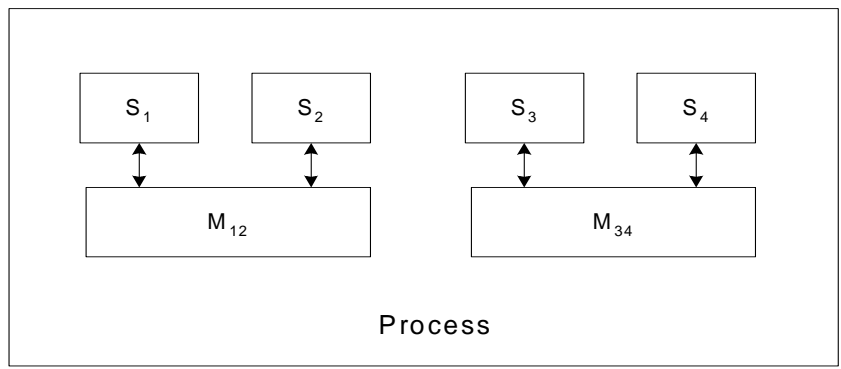

**Figure 2**: Design goal of the Weaves framework. We need to support multiple solvers linked to mediators within a single process.

Before we attempt to design a programming model that can support the interaction shown in Figure 2, let us begin with the case depicted in Figure 3. Here, two PDE solvers are linked to a single mediator performing an interface relaxation.

---

[1] This is not a requirement of the Weaves framework. Technically, all the solvers could implement different algorithms. However, the interesting cases arise when there are multiple instantiations of a given PDE algorithm.

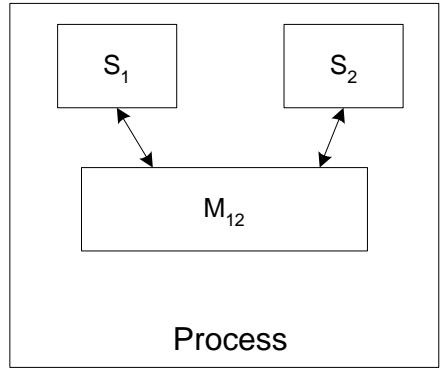

**Figure 3:** A simple compositional model with two PDE solvers linked to a single mediator. This composition addresses a composite PDE problem involving two domains.

In order to model the composition shown in Figure 3, let us start with a simple *process per solver* model as shown in Figure 4a. This model allows a single PDE solver application to be linked to a mediator. The external references of the PDE solver application will be bound to the mediator. Multiple such processes can be executed to simulate a network of collaborating solvers. This simple approach has several problems. The first and most basic problem is that we cannot link multiple solver applications to a *single shared* mediator without modifying the application. The second major problem with the process model arises from scalability concerns. A large network of collaborating PDE solvers will involve tens to hundreds of interfaces, each of which is a process. Inter-process context switch time will be a major bottleneck.

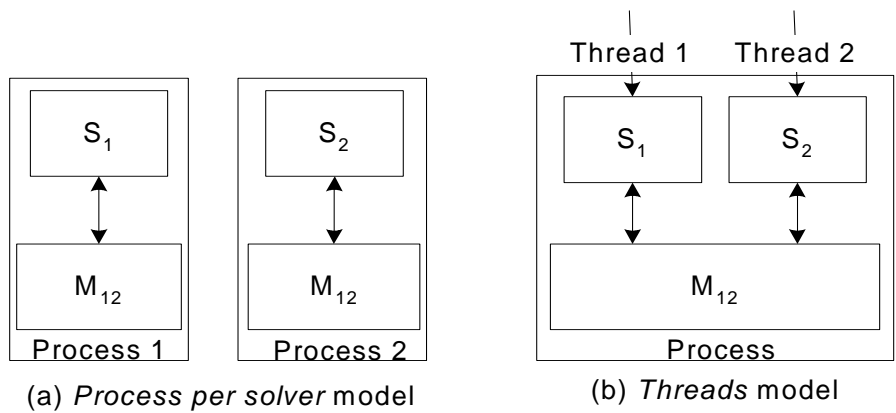

**Figure 4**: The composition shown in Figure 3 modeled under (a) process per solver model and (b) threads model. Neither of these models can achieve the composition shown in Figure 3.

To address the above concerns, let us build a new model based on threads as shown in Figure 4b. In the threads model, the composition shown in Figure 3 can be achieved by modifying the PDE solver application source to create two threads, or inserting a piece of stub code that creates two threads, with the start function of each thread set to the entry point of the solver application. As before, the mediator application is linked to the PDE solver.

The major problem with this approach arises from updates to global variables - in particular, the PDE solver may not be thread-safe. Since threads share global variables, a solver thread modifying a global variable will inadvertently change the state of the other – unrelated - solver thread causing erroneous behavior. Ideally, we need two copies of all the global variables used in the PDE solver. In programs that are explicitly threaded in design, sharing of global state is intentional. In our case, this *sharing is neither intentional nor necessarily desirable*. On the other hand, to create a shared mediator, we need to share global state within the mediator between the threads running through it.

The threads example illustrates the crux of our problem – *conflicting needs – the need to **avoid sharing** global state between threads of the solver application and the need **for sharing** global state between the threads running through the mediator*. The intuition here is that we need a programming model that allows arbitrary sharing of global state. Such a model can subsume both the thread and process models, since it can allow both complete sharing of global state as in threads as well as no sharing as in processes This observation leads us to the first step towards Weaves DCE framework.

## 3.1 Defining the Framework

The major components of the Weaves programming framework are:
1. **Module**: A module is any object file or collection of object files defined by the user. Modules have:
2. A **data context**, which is the global state of the module scoped within the object files of the module, and
3. A **code context,** which is the code contained within the object files that constitute the module. The code context may have multiple entry point and exit point functions.
4. **Bead**: A bead is an instantiation of a module. Multiple instantiations of a module have independent data contexts, but share the same code context.
5. **Weave**: A weave is a collection of data contexts belonging to beads of different modules. The definition of a weave forms the core of the Weaves framework. Traditionally, a process has a single name space mapped to a single address space. Weaves allow users to define multiple namespaces within a single address space, with user-defined control over the creation of a namespace.
6. **String**: A string is a thread of execution that operates within a single weave. Similar to the threads model, multiple strings may execute within a single weave. However, a single string cannot operate under multiple weaves. Intuitively, a string operates within a single namespace. Allowing a string to operate under multiple namespaces would violate the single valued nature of variables.
7. **Tapestry**: A tapestry is a set of weaves, which describes the structure of the composed application. The physical manifestation of a tapestry is a single process.

The above definitions have equivalents in object-oriented programming. A module is similar to a class and a bead - which is an instantiation of a module – is similar to an object. Tapestries are somewhat similar to object hierarchies. The major exception is that interaction between beads within a tapestry involves runtime binding. We chose to use our own terminology to (a) avoid overloading the semantics of well-known OOP terms and (b) avert the implication that the framework requires the use of an OOP language.

Strings are similar to threads in that (a) they can be dynamically instantiated and (b) they share the same copy of code. However, unlike threads, strings do not share global state. Each string has its own copy of global state. The main goal here is to avoid inadvertent sharing of state between unrelated instantiations of an algorithm, without having to modify the algorithm.

Since strings are an intra-process mechanism, we will illustrate their operation by comparing and contrasting them to threads. A thread's state consists of (a) an instruction pointer (**IP**), (b) a stack pointer and (c) copy of CPU registers. Each thread within a process has its own stack frame that maintains local variables and a series of activation records that describes the execution path traversed by the thread.

When a thread is created, the thread library creates a new stack frame and starts execution at the first instruction of the function specified by the thread instantiation call. When the thread scheduler needs to switch between threads, it saves the current IP, current stack frame, and the values in the CPU registers, switches to the state of the next thread, and starts execution at the IP contained in the thread state.

Strings involve an extension to the operation of threads. Similar to threads, each string has its own stack frame, which maintains local state. In addition, each string also has a copy of the global variables in an area called the *weave context frame*, the start of which is pointed to by a *weave context frame pointer*. A weave context defines the namespace of a string. This includes the global variables of all the beads traversed by a string. Note that some of the beads in a string may be shared between strings.

A string's state consists of (a) an instruction pointer, (b) a stack frame pointer, (c) copy of CPU registers and (d) a weave context frame pointer. When a string is created, the string creation call creates a stack frame and a weave context frame (if necessary) and copies the current state of the global variables into the weave context frame. The string creation call also associates a numerical identifier with the newly created string. Since creating a string involves copying its global variables, the string creation cost depends on the storage size of the global variables resulting in a higher creation cost than threads. We justify this cost by noting that it is a one time cost paid at program startup. Also, well-written applications are generally frugal in their use of globals, which mitigates the impact of the copy operation.

Similar to a thread scheduler, the string scheduler starts execution of the new string at the first instruction of a user-specified function. When the string library needs to switch between strings, it saves the current IP, current stack frame pointer, the values in the CPU registers, and the current weave context frame pointer, switches to the state of the next string and starts execution at the IP contained in the string state. The inter-string context switch cost is identical to threads.

Selective sharing of state in our framework operates at the level of individual beads. We illustrate the operation of selective sharing with the example shown in Figure 2 (also repeated in the Figure 5 below). The tapestry defines 4 weaves <Solver $S_1$, Mediator $M_{12}$>, <Solver $S_2$, Mediator $M_{12}$>, <Solver $S_3$, Mediator $M_{34}$> and <Solver $S_4$, Mediator $M_{34}$>, and 4 strings, with each string operating within a single weave. At run time, context switching between the strings automatically switches the namespace associated with the string, preserving the sharing specified in the tapestry.

Figure 5 depicts the design process in the weaves framework. The design process involves two entities: a *programmer* who implements the modules and a *composer*, who uses a graphical user interface to instantiate beads and define the various weaves and strings. The result of the GUI composition is a tapestry configuration file, which is used to load and execute the composed application. Each composed application also has a module called a *monitor* that is automatically linked with the composed application. In the process model, utilities like *ps* (in UNIX) can be used to query the run time of the process. The monitor provides a much more powerful IPC (Inter Process Communication) interface to such functionality. Utilities can query the monitor to determine the current tapestry, beads, strings, and weaves within a composed application.

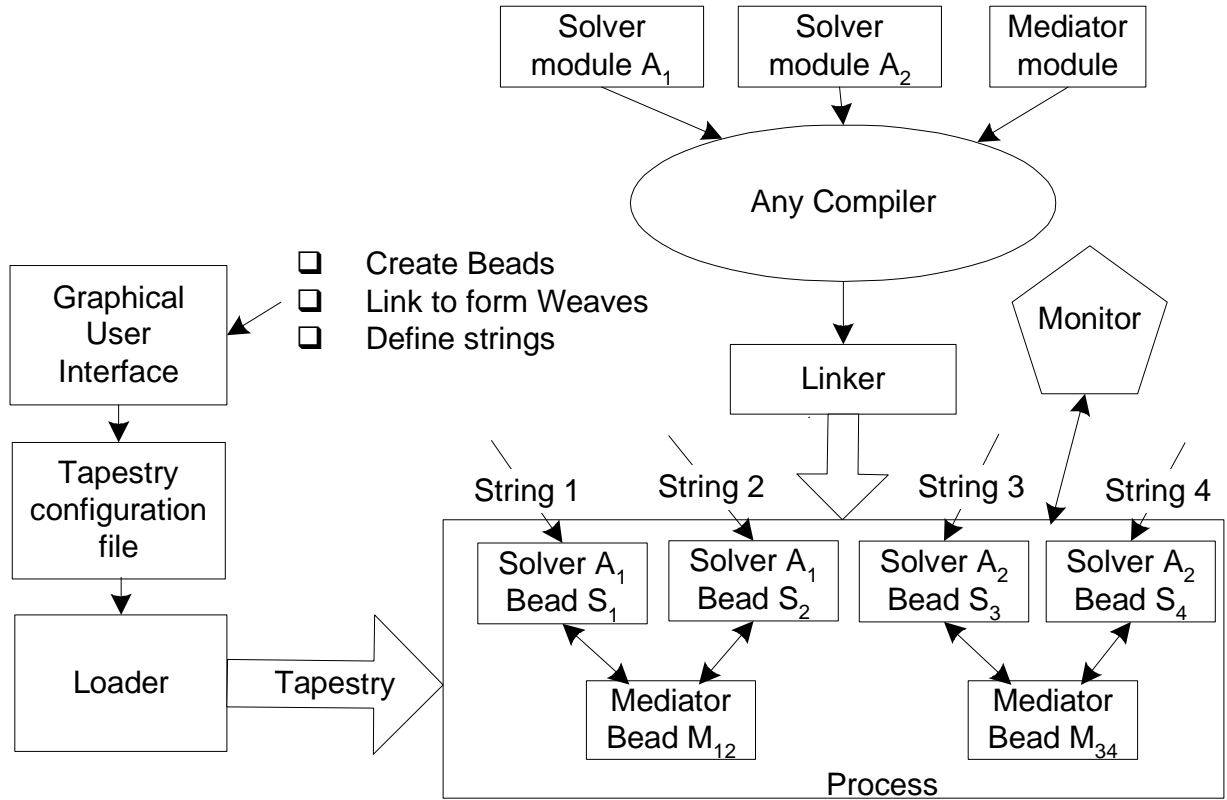

**Figure 5:** Interaction between the various components of the Weaves framework.

The tapestry generated by the GUI is not necessarily a static composition. The Weaves programming framework allows applications to rewire themselves on the fly in response to dynamic conditions. Two forms of dynamic application composition are supported in the framework. In the first form, if the requisite modules are already linked into the original tapestry, Weave-aware applications can modify their structure by creating new beads, defining weaves, and instantiating strings at run-time. For non-Weave aware applications, the interface exposed by the monitor can be used to modify the tapestry of a composed application. These modifications may be manually made by a user at the command line or can be automatically generated by an external *resource monitoring* agent.

In the second form of dynamic composition, new code modules can be inserted into a running application through a modified dynamic library interface. In this mode of operation, the dynamically inserted code is analyzed at run-time. Dynamically inserted modules can be used in the same manner as statically inserted modules. This interface provides the full capabilities of Weaves, including arbitrary namespaces and compositional capabilities, in a run-time compositional framework.

The astute reader may have observed that all the examples of composition presented thus far have an *inverted tree* structure, with sharing occurring at the lower layers. Intuitively, the inverted tree structure appears "natural." This is because it mirrors the sharing and flow of control relationships between processes in an operating system running on a single processor architecture. The resources within the operating system are at a lower layer and shared between the independent processes, which naturally leads to an inverted tree structure.

While the Weaves framework is capable of implementing regular tree structures and context switching between the various strings, from a usability perspective, scheduling in a regular tree structure requires application support. In an inverted tree structure, string scheduling decisions can be made independently of the application without impacting either usability or correctness. However, in a regular tree structure the same does not hold true. Consider this problem from the perspective of the application programmer. A call to a function in a regular tree structure would map to two or more independent data contexts. Choice of the correct data context can only be made by the programmer. The solution in the Weaves framework is to provide support functions that allow the programmer to (a) query the string for its identifier and (b) base the path of execution on the string identifier. This solution is similar to the SPMD programming model on distributed memory architectures, where the execution path is determined by the processor identifier [Darema, 2001].

Another issue in the Weaves framework arises from reentrancy concerns at the lower layers of an inverted tree. In particular, the problem arises when preemptive scheduling strategies cause reentrancy in beads shared at lower layers, when the codes are not reentrant. We solve this problem by organizing strings into equivalence classes, where each equivalence class contains strings that share beads. Preemptive scheduling switches between strings of different equivalence classes. If the preempted string has not traversed a shared bead, preemptive scheduling can also switch between strings of the same equivalence class. The details of the solution are beyond the scope of this article.

## *3.2 Tuple Spaces*

The notion of selective state sharing in the Weaves programming framework presents a very powerful mechanism for defining namespaces. Since the definition of a weave permits any set of beads to define a namespace, any composition that can be represented by a connected graph (or a set of independent graphs) can be realized by this framework. From an application's perspective, the definition and operation of distinct namespaces is transparent. This mechanism presents a powerful compositional framework for any procedural code.

The Weaves programming framework also supports the notion of shared tuple spaces. In the current definitions, distinct beads of the same module have different data contexts, i.e. data sharing occurs at the granularity of an entire module. To create a shared tuple space, we need fine grain control over the individual members of a data context.

In order to support shared tuple spaces, from the perspective of the framework, we need mechanisms to (a) define a shared tuple space and (b) to selectively share the members of the tuple space across multiple beads. To define a shared tuple space, application composers can use a graphical user interface to denote the members of the tuple space or code modules can use a syntactic notation to mark the members of the tuple space. This information is used at bead creation time to merge references to shared members of a tuple space.

# 4 Configuring Weaves for High Performance Scientific Computing

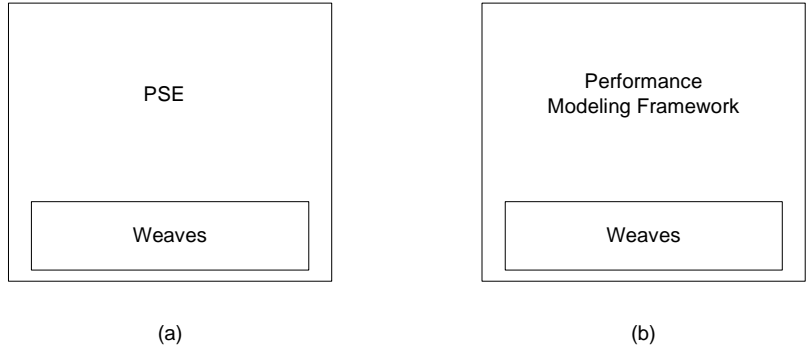

**Figure 6:** Relationship between Weaves and (a) a problem solving environment and (b) a performance modeling framework.

In any programming model of the kind presented here, it is instructive to examine the various configurations in which it can be used in a high performance scientific computing environment. Two basic choices are presented in Figure 6. The simpler configuration (Figure 6 (a)) involves invoking Weaves as part of a larger PSE; here, Weaves forms the underlying execution manager that exposes abstractions to configure and execute tapestries composed by the PSE. The PSE can also use the runtime adaptation capabilities of Weaves to dynamically modify and steer computations.

Figure 6 (b) shows Weaves operating within a larger performance modeling framework, such as POEMS [Adve et al., 2000]. Here, POEMS supplies the modeling capability needed to characterize performance and Weaves takes care of the low-level composition and direct code execution of unmodified scientific codes. Thus, systems such as POEMS can utilize Weaves as the computational substrate on which modeling is performed. POEMS has expressive representations (e.g., task graphs) that unify analytical modeling, simulation, and actual system execution paradigms [Adve and Sakellariou, 2000]. Task graphs help POEMS to address both model execution and model analysis goals. For instance, task graph compiler analysis has been shown to improve simulation capabilities by identifying opportunities for reducing (and removing) computational overhead [Deelman et al., 2001]. The results of such analyses can be communicated to Weaves either *a priori* via the tapestry configuration file, or dynamically at runtime, leading to more efficient direct code execution.

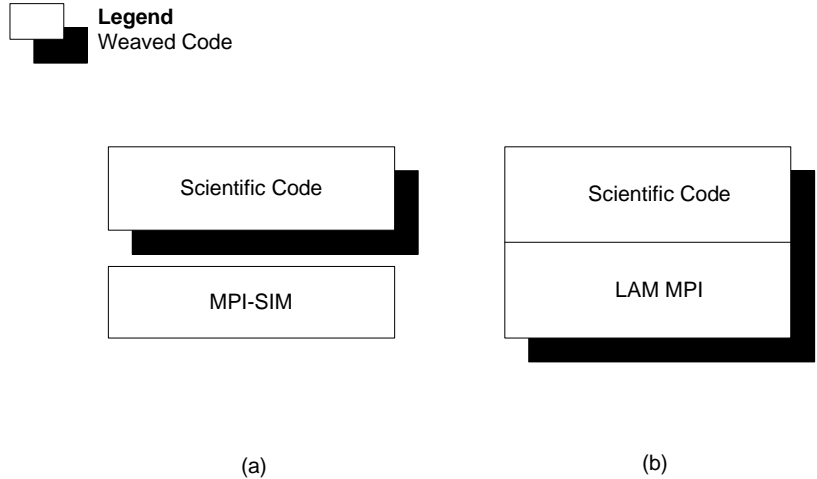

**Figure 7:** What can be weaved: (a) scientific codes running over MPI-SIM and (b) both scientific codes as well as the underlying MPI implementation.

Two final choices of configuration involve "opening up" the performance modeling framework to identify further opportunities for weaving. For instance, one of the primary functionalities supported by POEMS is the simulation of MPI-based codes using the MPI-SIM simulator [Prakash and Bagrodia, 1998]. MPI-SIM uses a multi-threaded architecture that simulates the complex inner workings of MPI. However, MPI-SIM assumes that the linked code base is thread-safe. While this assumption is met by codes such as Sweep3D [Adve et al., 2000], it may not hold in other cases.

To ameliorate this scenario, we propose the configurations of Figure 7. In Figure 7 (a), Weaves is used to present a thread-safe view of the application to MPI-SIM. Since MPI-SIM is by itself thread-safe, this configuration of Weaves enables a wider variety of applications to be simulated under MPI-SIM. This is an instance of synergy between Weaves as a programming paradigm and MPI-SIM as an emulation framework. The former factors out the non-thread-safe components of the application while the latter concentrates on the detailed modeling of the communication primitives.

Figure 7 (b) is the most radical of our configurations. It take the model presented in Figure 7 (a) a step further by weaving the MPI library itself (e.g., the LAM MPI implementations), linking it to the application and creating multiple instantiations similar to the process model. This configuration has the advantage of being able to conduct direct code execution of not just the application, but the application in the context of its original communication interface. The representation created is a true (native) virtual machine abstraction for MPI codes. It enables the study of the effects of different MPI implementations on performance of scientific codes. In this model, the only simulated entity is the underlying communication channel, not the operation of MPI.

## 5 Implementation and Evaluation

The core of the Weaves DCE framework is the abstraction of a weave, which allows an application composer to define arbitrary namespaces over a composed application. To implement the weave abstraction, we need a data structure that can efficiently capture the state separation and state recombination needs of the DCE framework.

Before we discuss the specifics, note that the goals of the DCE framework place additional constraints on the implementation of the weave abstraction. First, our transparency requirement states that the solution should be transparent to the application. Since, the application may be written in any programming language, the transparency requirement precludes modification to the source code to implement the namespace abstraction. Second, from a scalability perspective, the implementation should be efficient. In particular, we need to minimize context switch time between the various namespaces defined in the composed application.

To meet the transparency requirement, the implementation of the namespace abstraction works by analyzing the Executable and Linking Format (**ELF**) object files produced by any compiler. ELF is a public domain file format used to represent both object code as well as the final executable on most UNIX systems. Our current prototype is implemented on the Linux operating system running on Intel x86 architectures. Since the implementation only depends on the ELF file format, it can be easily ported to other operating systems/architectures. Furthermore, we anecdotally note that the features of the ELF file format used by our implementation are common to object file formats. Hence, it should be possible to extend the prototype to support other object file formats as well.

The ELF file format uses the Global Offset Table (**GOT**) data structure to access global state in an application. The GOT data structure maintains an array of *pointers* (instead of data values), with each pointer referring to a global data variable. To access data, applications first index into the GOT data structure to get a pointer to the data and then use the pointer (and possibly an offset) to retrieve the data value. The number of entries in the GOT structure is proportional to the number of variables and is independent of the size of each variable. For instance, an array variable has a single entry in the GOT structure. The observation here is that the GOT defines the namespace of the application. Typically, a program contains a single GOT structure reflecting the single namespace within an application. However, by appropriately defining multiple GOT structures, it should be possible to create multiple namespaces within a single ELF executable.

The problem with the basic GOT structure is that compilers hardcode the base address of the GOT structure and the index into the GOT at compile time. To implement multiple namespaces, we need to create multiple GOT structures and, at runtime, copy them over to the fixed base address generated by the compiler. This operation is expensive since its cost is proportional to the number of global variables, which can potentially be large.

Instead, we note that compilers produce relocatable code (for instance, the command line option –fPIC on the gcc family of compilers) to support dynamic libraries. In relocatable codes, the base of the GOT structure is pointed to by a base register. All indexed accesses into the GOT are made with reference to the current value of the base register. The use of relocatable object code and indexed access to the GOT forms the basis of our implementation.

To implement the weave abstraction, we create a new GOT structure for each distinct weave in the composed application. To implement state separation between beads belonging to different weaves, we first create copies of the data and point the GOT entries in the weaves to the distinct copies of the data. To enable state recombination between weaves sharing a bead, we set the pointer in the GOT entries in the different weaves to point to the same data value. The double indexed nature of the GOT structure enables state separation/recombination at the resolution of a single data variable, which can be used to implement arbitrary data sharing at both the tuple space and module levels.

To implement the string abstraction, note that a string is really a thread operating under a user specified namespace. Since we have a mechanism to create the namespace, context switching between strings involves context switching the thread state and switching the namespace. What we need here is an efficient mechanism for switching namespaces.

To switch namespaces, we note that the GOT structure is accessed through a base register (%ebx in our current implementation). Hence, context switching between namespaces merely involves changing the base register to point to a different GOT structure, a single instruction move operation. Furthermore, we note that we do not even have to pay the cost of this instruction. The base register is automatically saved and restored by the thread context switch routine. If we initialize the base register to point to different GOT structures before thread creation, the thread context switching routines automatically context switch between weaves. This results in a weave context switch time that is identical to thread context switch time. Our current implementation of the Weaves framework works over both POSIX Threads (pthreads) as well as the GNU Portable Threads (Pth) thread libraries.

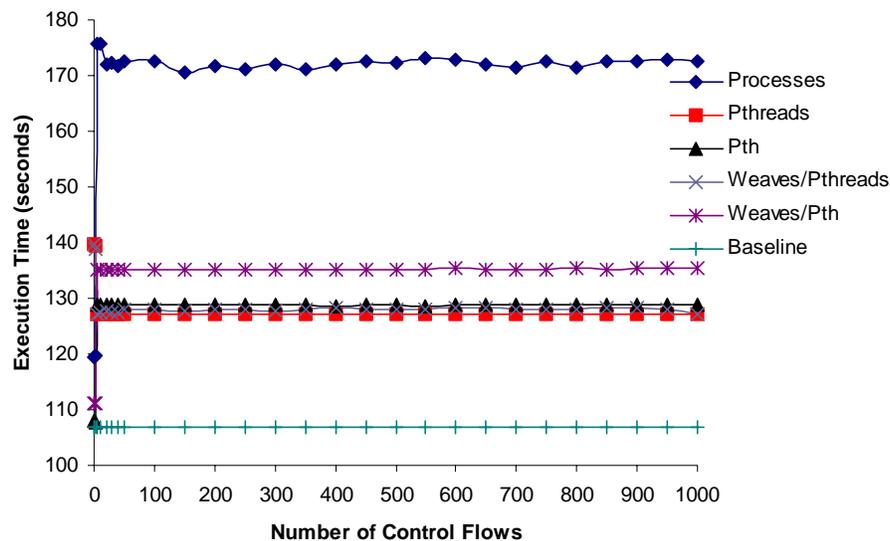

**Figure 8:** Comparison of inter-flow context switch time in the threads, processes, and weaves programming models. The baseline single process application implements a calibrated delay loop of 107 seconds.

We ran a series of experiments to compare the context switch time under the threads, processes and weaves programming models. In this experiment, we created a baseline application that implements a calibrated delay loop (busy wait). We then implemented threads-, processes-, and weaves- versions of the application. In each of these versions, there are $n$ independent flows of control over the same code, where each flow of control executes a calibrated delay loop, which does $1/n^{th}$ the work of the baseline application. We then measure the total time taken to execute the application under each of these models. Since each of the control flows does $1/n^{th}$ of the work and there are $n$ flows, the total time taken should the same as the baseline calibrated delay loop case, except for an additional context switching cost.

Figure 8 shows the results of the experiment on a single processor AMD Athlon™ workstation running the Linux operating system. The results show the run time for five cases: (a) baseline calibrated delay loop, (b) pthreads threads library, (c) Pth threads library, (d) processes, (e) Weaves over pthreads, and (f) Weaves over Pth. The results clearly show that the weaved implementations are significantly faster than processes, even in this simple case, where the copy-on-write semantics of the *fork()* call are very effective. Furthermore, the run time of weaved implementation of pthreads is very close to the base run time of pthreads alone. The marginal variation in runtime is due to the slightly higher weave creation cost, which is included in the run time. Also, the pthreads implementation is relatively efficient, since the Linux kernel includes operating system support for it.

However, in the case of Pth, the run time of the weaved implemented is higher than the base Pth case. This increase in runtime is because unlike pthreads, Pth is a user-level library and hence suffers from timer inaccuracies inherent in user-level library implementation. As an aside, we mention that we noticed several discrepancies in the Linux operating system scheduler. The base processes and pthreads implementations showed far less variation on the SGI IRIX™ operating system.

## 5.1 Weaving Sweep3D

To test the capabilities of the Weaves framework on real-world scientific codes, we weaved the Sweep3D application [Adve et al., 2000]. To support the message passing primitives used by Sweep3D, we created a simple threaded MPI emulator, which implements only the **nine** MPI primitives used by Sweep3D. To ensure correctness, the MPI emulator implementation follows the guidelines set forth in the MPI specification. Our MPI emulator is intended as a test prototype and is neither as comprehensive nor as capable as MPI-SIM.

In the weaved implementation, we create *n* distinct virtual machines, each of which executes an independent instantiation of the Sweep3D application. To do this, we create *n* distinct Sweep3D beads and *n* weaves, where each weave has a distinct Sweep3D bead and a shared emulator bead. Each weave also has a single string associated with it. The *n* distinct virtual machines run on a single processor workstation.

We compared the performance of our single processor weaved implementation of Sweep3D against measured values from real runs for up to 150 processors. Measurements for the real runs were made on our 200 processor cluster *Anantham*. Each node of *Anantham* has a 1GHz AMD Athlon ™ processor, with 1GB RAM. The nodes are interconnected over a switched Myrinet communication fabric, which provides 4Gbps of network bandwidth per node. Since the Sweep3D application performs its own timing measurements, we compared the timing numbers (CPU Time) of the weaved version of Sweep3D with the measurements from actual runs. The two input files (50x50x50 and 150x150x150 decompositions) provided in the Sweep3D distribution were used to drive the Sweep3D application.

For upto 150 processors, the timing results from the weaved implementation and the actual runs were consistent to within 0.2%. Furthermore, we tested the weaved version of Sweep3D with over 1000 weaves on a single processor. The variation in the timing results between multiple runs was within 0.2%. This clearly shows that even at high levels of scalability (over 1000 weaves/processor) context switch time does not impact the efficacy of the Weaves DCE framework.

## 6 Discussion

We have shown how Weaves supports a novel compositional framework for combining and executing unmodified high performance scientific codes. In contrast to traditional component-based models, Weaves takes a low-level, namespace view to the problem of composition. This enables the migration of the vast majority of legacy scientific codes to a component-based framework, with all the associated benefits.

Other functionality enabled by Weaves, but which we haven't studied in detail here, are transparent checkpointing and runtime load balancing through string migration. We plan to elaborate on these in a future paper. This will support the extension of the Weaves framework to applications requiring model-based, adaptive control (e.g., those studied in [Adve et al., 2002]).